\documentclass[conference]{IEEEtran}
\IEEEoverridecommandlockouts
\usepackage{cite}
\usepackage{amsmath,amssymb,amsfonts}
\usepackage{graphicx}
\usepackage{textcomp}
\usepackage{xcolor}
\usepackage{multirow}
\usepackage{epsfig}
\usepackage{algpseudocode}
\usepackage{algorithm} 

\def\BibTeX{{\rm B\kern-.05em{\sc i\kern-.025em b}\kern-.08em T\kern-.1667em\lower.7ex\hbox{E}\kern-.125emX}}

\newcommand{\linebreakand}{%
  \end{@IEEEauthorhalign}
  \hfill\mbox{}\par
  \mbox{}\hfill\begin{@IEEEauthorhalign}
}

\begin{document}

\title{TwinPot: Digital Twin-assisted Honeypot for Cyber-Secure Smart Seaports\\
}

\author{
\IEEEauthorblockN{
Yagmur Yigit\IEEEauthorrefmark{1}
Omer Kemal Kinaci\IEEEauthorrefmark{2}
Trung Q. Duong \IEEEauthorrefmark{3}, and 
Berk Canberk \IEEEauthorrefmark{4} 
}
\IEEEauthorblockA{\IEEEauthorrefmark{1} Department of Computer Engineering, Istanbul Technical University, Turkey \\
\IEEEauthorrefmark{2} Department of Shipbuilding and Ocean Engineering, Istanbul Technical University, Turkey \\
\IEEEauthorrefmark{3} School of Electronics, Electrical Engineering and Computer Science, Queen's University Belfast\\
\IEEEauthorrefmark{4} School of Computing, Engineering and The Build Environment, Edinburgh Napier University, United Kingdom \\
Email: \{yigity20, kinacio\}@itu.edu.tr, trung.q.duong@qub.ac.uk, B.Canberk@napier.ac.uk}
}

\maketitle

\begin{abstract}
The idea of next-generation ports has become more apparent in the last ten years - in response to the challenge posed by the rising demand for efficiency and the ever-increasing volume of goods. In this new era of intelligent infrastructure and facilities, it is evident that cyber-security has recently received the most significant attention from the seaport and maritime authorities, and it is a primary concern on the agenda of most ports. Traditional security solutions like firewalls and antivirus software can be applied to safeguard IoT and Cyber-Physical Systems (CPS) from harmful entities. Nevertheless, security researchers can only watch, examine, and learn about the behaviors of attackers if these solutions operate more transparently. Herein, honeypots are potential solutions since they offer valuable information about the attackers. Honeypots can be virtual or physical. Virtual honeypots must be more realistic to entice attackers, necessitating better high-fidelity. To this end, Digital Twin (DT) technology can be employed to increase the complexity and simulation fidelity of the honeypots. Seaports can be attacked from both their existing devices and external devices at the same time. Existing intrusion detection mechanisms are insufficient to detect external attacks; therefore, the current systems cannot handle attacks at the desired level. DT and honeypot technologies can be used together to tackle them. Consequently, we suggest a DT-assisted honeypot, called TwinPot, for external attacks in smart seaports. Moreover, we propose an intelligent attack detection mechanism to handle different attack types using DT technology for internal attacks. Finally, we build an extensive smart seaport dataset for internal and external attacks using the MANSIM tool and two existing datasets to test the performance of our system. We show that under both simultaneous internal and external attacks on the system, our solution successfully detects internal and external attacks.
\end{abstract}


\section{Introduction}
\label{sec:intro}

In parallel with the rapid development of technologies like the Internet of Things (IoT) and Artificial Intelligence (AI), the transportation sector has also started to digitize itself quickly.
Businesses nowadays are pushing more than ever to increase productivity. Along with this push is the industry-wide, ever-growing desire to curb negative environmental impacts. 
Global transportation is one of the businesses that increasingly demands efficiency in operations. The maritime industry handles 90\% of all goods transported around the world \cite{mdpi22}. Maritime transportation and national economies are heavily dependent on seaports and their infrastructure. In this manner, digitalization is essential in delivering measurable savings and operational efficiency for seaports and their facilities.

The idea of next-generation ports has become more apparent in the last ten years in response to the challenge posed by the rising demand for efficiency and the ever-increasing volume of goods. However, this evolution was only possible with further developing technologies like the IoT and AI. Next-generation ports, also known as ``smart ports'',  focus on using intelligent solutions in seaport environments. As reported by Markets and Markets' Tech Report, the smart port market is forecast to grow exponentially at a rate of 24.3\%, to reach a total value of \$5.7 billion by 2027, from an estimated  \$1.9 billion in 2022 \cite{Market}. 
A smart port utilizes innovative technologies to boost productivity and safety while reducing expenses. Smart ports today use wired networks for connectivity. In contrast, future smart ports will use wireless technology such as Wi-Fi, Long Range Radio (LoRa), and 6G wireless network for connection \cite{ericsson}.  

\begin{figure*}[t]
    \centering
    \includegraphics[width=5.6in]{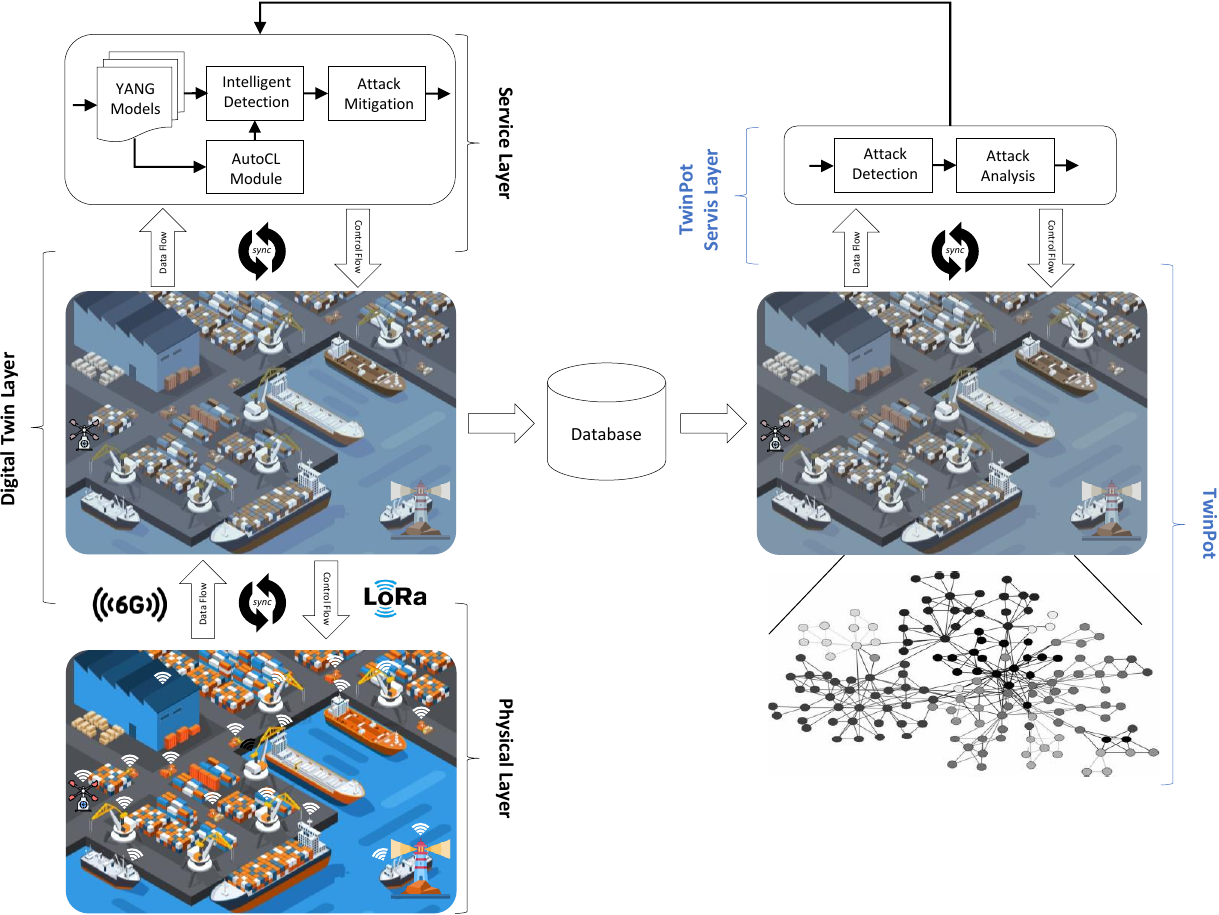}
    \caption{The proposed system architecture.}
    \label{fig:system}
\end{figure*}

In this new era of intelligent infrastructure and facilities, it is obvious that cyber-security has recently received the most significant attention from the seaport and maritime authorities, and it is a primary concern on the agenda of most ports. Between February 2020 and May 2020, the number of cyberattacks against the marine sector worldwide increased fourfold, according to the International Association of Ports and Harbors (IAPH). When looking at a longer timescale, attacks against the operational systems soared by an astonishing 900\% from 2017 to 2020.
%
%
For instance, the May 2020 cyberattack on Iran's Port of Shahid Rajaee caused system shutdown, traffic congestion, and halted cargo operations, leading to manual operations and reduced efficiency \cite{IAPH}.

Traditional security solutions like firewalls and antivirus software can be applied to safeguard IoT and Cyber-Physical Systems (CPS) from harmful entities. Security researchers can only watch, examine, and learn about the behaviors of attackers if these solutions operate more transparently. Herein, honeypots are potential solutions since they offer valuable information about the attackers. A security tool that sets up a virtual trap to entice attackers, used with the intent of being attacked and potentially compromised, is called a honeypot \cite{HoneySurvey}. 
There are both virtual and physical honeypots. Higher fidelity for virtual honeypots is required because they must be more realistic to tempt attackers.
To this end, the honeypots' complexity and simulation fidelity can be increased using Digital Twin (DT) technology. Thus, honeypots can become more attractive to attackers.
Seaports can be attacked from both their existing devices and external devices at the same time. Existing intrusion detection mechanisms are not sufficient to detect external attacks. Therefore, the current systems cannot handle attacks at the desired level.
DT and honeypot technologies can be used together to tackle them. Therefore, we suggest a DT-assisted honeypot, called TwinPot, for external attacks in smart seaports. TwinPot provides a better analysis of attacks and contributes to improving the cybersecurity mechanism of smart seaports.
Furthermore, we propose an intelligent attack detection mechanism to handle different attack types using DT technology for the current entities in seaports.

Our main contributions are as follows:
\begin{itemize}
    \item By utilizing DT and honeypot technologies for smart seaports, we propose a TwinPot to analyze the behaviors of attackers for external entity attacks in the seaport and improve the security mechanisms.  
    \item We suggest a network-aware intelligent attack detection mechanism to tackle the attacks through current seaport entities taking advantage of DT for smart seaports. We implement an Automated Classification Method (AutoCM) to handle different attack types.
\end{itemize}

The remainder of the paper is structured as follows. A review of related studies in the literature is given in Section~\ref{sec:relWork}. Section~\ref{sec:proposed} and Section~\ref{sec:perfEval}, respectively, explain the proposed solution and the performance evaluation. Finally, we conclude the paper in Section~\ref{sec:conc}.

\section{Related Work}
\label{sec:relWork}

Security is an essential topic for the maritime sector. F. Akpan \textit{et al.} emphasized security problems and challenges in the maritime sector \cite{maritime22}. They provided potential vulnerabilities of the ship systems and their possible solutions. 
Another work highlighted the importance of protecting critical infrastructures while facilitating new technologies in the new ``4.0 world'' \cite{elsevier21}. This work highlighted that raising connectivity in smart ports creates a unique ecosystem and threats.
G. D’Amico \textit{et al.} provided a systematic literature analysis that identifies the recurrent themes in logistical projects for intelligent and sustainable port cities \cite{portcities21}. Their proposed multidimensional framework integrates existing enabling domains for sustainable logistics, emphasizing the need for sustainable and intelligent logistics with digital technologies for data analysis and port cities.
%
Another work provided a summary of IoT systems security challenges of seaports \cite{mipro}. They gave potential weaknesses of IoT systems when utilized in seaports and their advantages, with a summary of current IoT-enabled seaports.

DT, another innovative technology like IoT, has started to be used in seaport environments, and there are currently a few works in this domain.
H. Li \textit{et al.} proposed a DT framework for next-generation ports \cite{wsc21}. Additionally, they asserted that SingaPort.NET is a collection of DT created by the suggested framework to address the simulation, optimization, and numerous modeling issues encountered by the maritime logistics sector.
R. Du \textit{et al.} gave the DT-powered potential 5G smart port use cases \cite{ericsson}. They predicted that all connections would be wireless in future innovative ports.
Another work proposed a DT framework to address communication and synchronization problem using IPv6 infrastructure and improves the accuracy of net-zero waste services \cite{T6conf}.
An innovative technique for using a DT of inland rivers based on 3D video fusion was put forth in another study \cite{access21}. The findings demonstrated that 3D video paired with DT improves river monitoring and emergency response by reducing the director's burden and enhancing effectiveness.

Although honeypots assist in detecting malicious requests and gathering harmful samples, they are rarely employed in IoT environments.
J. Franco \textit{et al.} gave a taxonomy and an exhaustive analysis of current honeypot studies for IoT, Industrial IoT (IIoT), and CPS \cite{HoneySurvey}. They outlined the open research issues that upcoming honeypot studies could tackle. This work pointed out that more research is needed since limited honeypot research for protecting innovative environments like ports and buildings from ransomware attacks.
%
W. Zhang \textit{et al.} proposed a hybrid honeynet that finds and records malicious behaviors and samples from the IoT network \cite{honeyIoT20}. Any physical node in the honeynet can receive commands and files from the honeynet control center at the same time. They put the suggested approach into place and effectively stopped numerous unknown harmful attacks.
B. Wang \textit{et al.} presented a honeypot framework for the IoT that consists of high and low interaction components \cite{IoTCMal}. After examining the homology of malware binaries collected by the suggested framework, they found types controlled by at least eleven attack groups.
None of the above studies specifically addressed attack detection and behavioral analysis in combination with honeypot and DT technologies for the intelligent seaports domain.

\section{Proposed Solution}
\label{sec:proposed}

We have two networks in our system: DT and TwinPot. The DT network exists with the physical, DT, and service layers. TwinPot and its service layer form the TwinPot network. The proposed system architecture can be seen in Fig.~\ref{fig:system}.
We divide the seaport network traffic into internal and external traffic in the proposed system. The traffic from the existing assets of the seaport is what we call internal traffic. We refer to external traffic as traffic from entities other than existing entities in the seaport, such as IoT devices on inbound cargo ships. We propose an intelligent detection technique to detect attacks that can be made from the existing assets of the seaport.
We duplicate the entities with their relation in DT to TwinPot. Thus, TwinPot's entities keep the interest of attackers from the main assets. After detecting an attack in TwinPot, behavioral analysis is made in the TwinPot service layer. This analysis is sent to the DT service layer to improve the real-system attack detection mechanism. If no attack is detected, incoming traffic is directed to the DT service layer.

\subsection{TwinPot}

We present TwinPot to provide complex and high-fidelity entities in honeypots. DT entities with their relation are duplicated to TwinPot. We use graph representation in TwinPot as in DT. We take data from the DT layer to a database periodically. We developed an algorithm to parse data automatically, taking the DT layer and removing critical data using YANG models. Then, the parsing data is sent to entities of TwinPot. We periodically send data to give a real network sense from the database to TwinPot. As a result, the TwinPot entities deter external attackers from targeting the real entities.

When an attack is detected in TwinPot, the system makes a behavioral analysis of the attack and defines its characteristics. After that, it sends a report to the DT service layer. This report finds and improves the current attack defense mechanism's drawbacks. On the other hand, normal external traffic transmits to the DT service layer.

\subsection{Intelligent Attack Detection Mechanism}
Our intelligent attack detection system works on internal seaport entities. Three layers build up the system: the physical layer, which contains physical entities; the digital twin layer, which contains digital twins of those same physical entities; and the service layer, which enables the attack detection mechanism. Algorithm~\ref{alg:detection} clarifies the pseudocode of the intelligent detection mechanism.
Our detection mechanism works on an entity basis and is as follows:
\begin{itemize}
    \item Twins of physical entities are established.
    \item Synchronization is built between the physical layer and the DT layer.
    \item Twins gather all data from physical entities in the physical layer using a networked sensor-based data acquisition system.
    \item Then, data is transferred to the service layer using YANG models.
    \item In the service layer, the system checks for any attacks according to the best classification method.
    \item If the system detects an attack, the attack mitigation module mitigates the attack and informs the system admin.
    \item If the system does not detect an attack, the reliability of the classification method is examined.
    \item If the reliability of the classification method is not less than its threshold, the system resumes operating with the existing model. 
    \item The features of the AutoCM are changed if the classification method's reliability falls below a predetermined threshold.
    \item AutoCM finds the best classification method between ten classification techniques for the system.
    \item After that, the AutoCM updates the system classification method as the best technique.
\end{itemize}

\begin{algorithm}[t]
\caption{the Intelligent Detection Mechanism}
\label{alg:detection}
\begin{algorithmic}[1]
\Require  Sensor data
\Ensure Attack security information and model reliability
\Procedure{\textit{Detection}}{}
\State Checking for any attacks on the system
\If {attack detected}
\State Run attack mitigation module
\State Inform system admin
\Else
\State Checking classification model reliability
\If {reliability not less than threshold}
\State Continue with the existing model
\Else
\State Change features of AutoCM
\State Finding the best method for the system
\State Updating system classification method
\EndIf
\EndIf
\EndProcedure

\end{algorithmic}
\end{algorithm}

\subsubsection{AutoCM Module}
Each classification approach has different results for different attack types. Therefore, we propose AutoCM to handle different attack types in smart maritime ports. AutoCM includes ten classification techniques: Naive Bayes (NB), K-Nearest Neighbors (KNN), Convolutional Neural Network (CNN), Decision Tree (DT), Recurrent Neural networks (RNN), Logistic Regression (LR), Long Short Term Memory (LSTM), Support Vector Machine (SVM), Multilayer Perceptron (MLP), and Random Forest (RF). 

According to how it operates, the offered AutoCM is:
\begin{itemize}
    \item With YANG models, a thousand features are imported into AutoCM.
    \item Since the taken data is unlabelled, we used the labeling algorithm to label data, and a baseline dataset as in our previous study \cite{CoreNet}.
    \item Data not labeled is then labeled through the labeling algorithm. It adds one thousand samples from the baseline dataset.
    \item Ten classification algorithms train and test their models using two thousand labeled data samples.
    \item After that, the final method algorithm determines the system's optimal classification method.
\end{itemize}

The metrics of the models are transmitted to the final method algorithm after training and testing. This algorithm determines the most effective classification technique.
Then, the system classification method is updated according to the best technique.

\subsubsection{Final Method Algorithm}
The final method algorithm regards false negative (FN), the value that occurs when the model forecasts the negative class inaccurately, false positive (FP), the value that occurs when the model mispredicts the positive class, and determination time. We utilize the following optimization objective in this algorithm:
\begin{equation}
   arg\,max \:(\alpha_i \lambda_i + \beta_i \nu_i), \:\: i \in [1, 10] 
\end{equation}
In equation 1, \emph{\(\lambda_i\)} is weighted sum of precision and recall of \emph{\(i_{th}\)} method, and \emph{\(\nu_i\)} is determination time of \emph{\(i_{th}\)} classification method. Among the ten techniques, we find the best classification method that has the maximum \emph{\(\lambda_i\)} and optimal \emph{\(\nu_i\)} values. We calculate \emph{\(\lambda_i\)} according to the following formula:

\begin{equation}
   \lambda_i = (0.6) \frac{TP}{TP+FN}  +  (0.4) \frac{TP}{TP+FP} , \:\: i \in [1, 10]
\end{equation}

In Equation 2, \emph{\(TP\)} presents the true positive, the value that occurs when the model correctly identifies the positive class, of \emph{\(i_{th}\)} classification method. The attack determination time is another critical metric for attack detection systems. Equation 3 depicts the determination time formula, in which finishing time is \emph{(\(t_{end}\))} and starting time is \emph{(\(t_{start}\))} of \emph{\(i_{th}\)} classification method.

\begin{equation}
    \nu_i = t^{end}_{i}\:-\:t^{start}_{i}, \:\: i \in [1, 10] 
\end{equation}

Furthermore, we defined the reliability of the classification method below the formula:

\begin{equation}
    \gamma = \{1 - \frac{FN}{TP+FN}\}
\end{equation}
In Equation 4, \emph{\(\gamma\)} stands for the reliability of the classification method. Since FN is a significant metric for data division, we define the method's reliability based on FN. 

\begin{figure*}[t]
    \centering
    \includegraphics[width=5.5in]{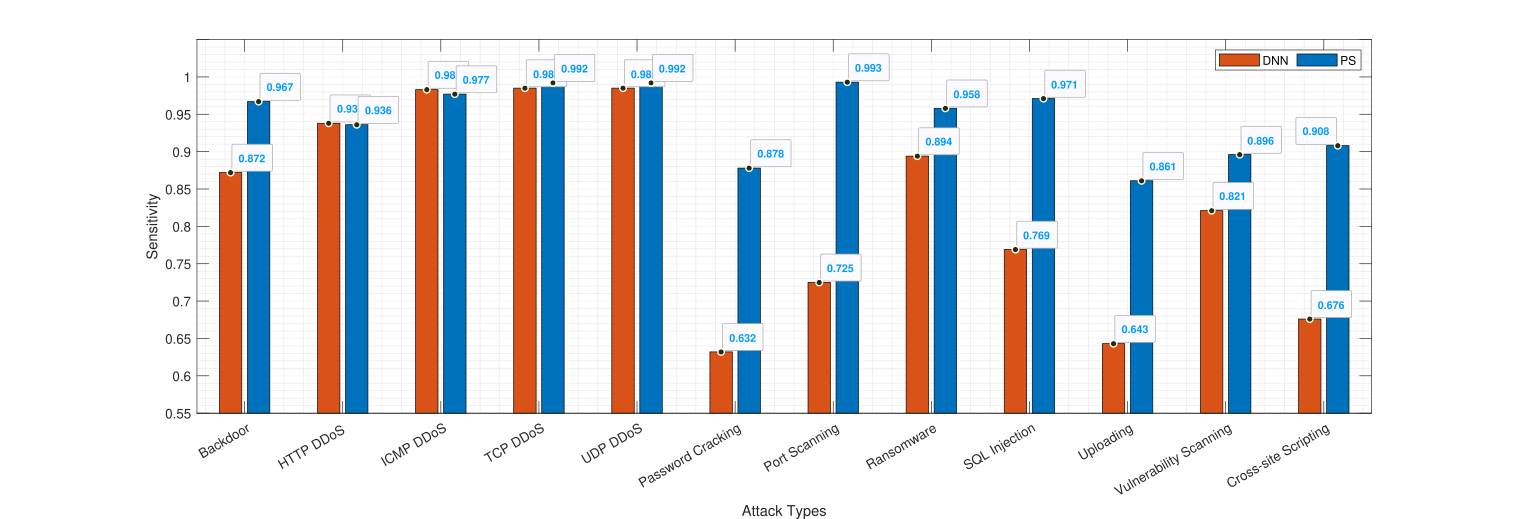}
    \caption{The performance comparison of the internal attacks.}
    \label{fig:internal}
\end{figure*}
\begin{figure*}[t]
    \centering
    \includegraphics[width=5.5in]{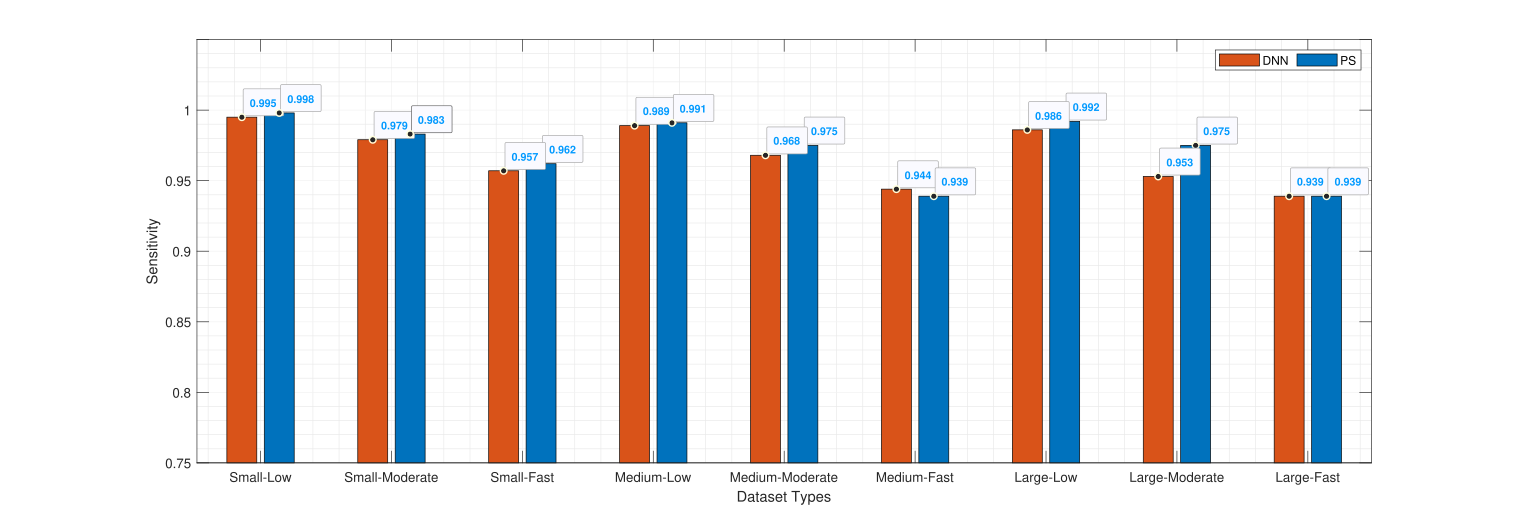}
    \caption{The performance comparison of the external attacks.}
    \label{fig:external}
\end{figure*}

\section{Performance Evaluation}
\label{sec:perfEval}

Twin graphs of the physical items are provided in this study using the Microsoft Azure DT (ADT) platform, which we employ as a service tool \cite{Azure}. ADT's features include the open modeling language, input and output plugins, DT Definition Language (DTDL), live representation, and more. 
We used MANSIM v2.01 (Maneuvering Simulation Laboratory), a maritime simulation code that generates time-dependent ship motion data for surface vessels to generate maritime data \cite{SUKAS2019a}. It includes 3DOF and 4DOF simulations for ships having single-screw/single-rudder and twin-screw/twin-rudder \cite{KINACI2021}. The code is freeware developed by one of the authors of this study and available online \cite{Mansim}. 
The code generally requires inputs such as ship particulars and environmental properties, hydrodynamic coefficients of the hull, propeller and rudder parameters, and solver parameters. The first three directly deal with the ship, its actuators, and the surrounding environment, while the last one is for determining the details of the time-based simulation. The code also allows controlled motions; an additional ``control parameters'' input exists in this case. Depending on the selection of the ``standard maneuvering test'' or ``free maneuver'' mode, the propeller rotation rate and the rudder angle might be input or output, respectively. Having identified these inputs, MANSIM returns the ship’s accelerations and velocities (translational and rotational) at each time step. The orientation of the ship and its location are also generated. We used these parameters to represent ship assets in the seaport. 
We also used two more datasets to represent IoT assets in the seaport and to analyze how well our solution detected attacks.
The Edge-IIoTset dataset is the first one, which includes fourteen attacks on connectivity protocols and was produced recently for the IoT and IIoT \cite{EdgeIIoT}. We used the data in this dataset as internal traffic for the system. 
We used the SDSN dataset produced in our previous study for external traffic \cite{SDSN}. It contains nine different DDoS UDP attack datasets that we created from heterogeneous traffic speed and variable traffic payload criteria.
Moreover, we compare our proposed solution (PS) with the Deep Neural Network (DNN) algorithm since it performs best in the work of M. A. Ferrag \textit{et. al} \cite{AccessIIoTset}.

\begin{figure}[t]
    \centering
    \includegraphics[width=2in]{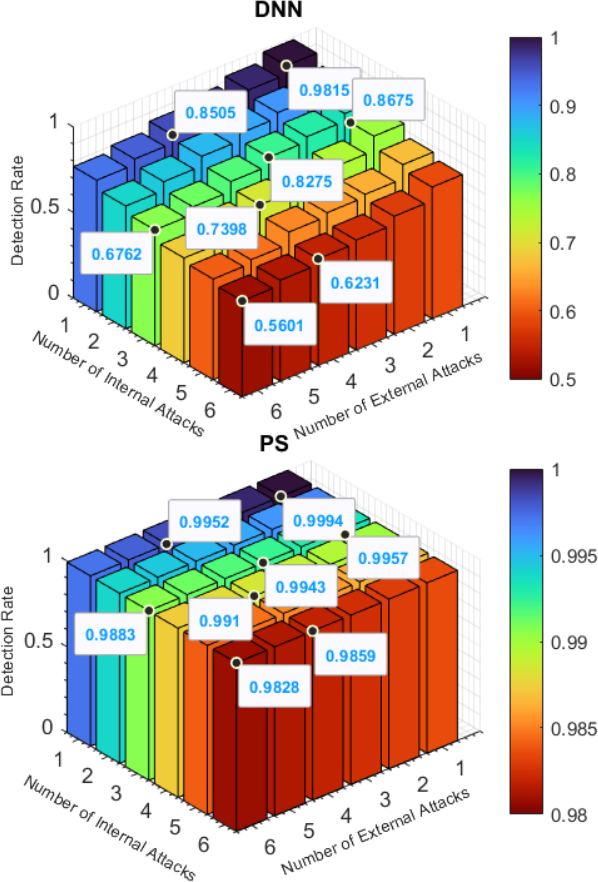}
    \caption{The detection rate comparison of simultaneous attacks.}
    \label{fig:detrate}
\end{figure}

We used five thousand samples of twelve attack traffic to test internal attacks and fifteen thousand samples of normal traffic in the Edge-IIoTset. We combined it with the twenty thousand data samples produced using the MANSIM tool to create an extensive smart seaport dataset for internal attacks. As a performance metric, we used sensitivity, the proportion of samples correctly identified as attacks to all samples that should have been attacked.
The performance results of our solution can be seen in Fig.~\ref{fig:internal} for the internal attacks.
We used two thousand samples of attack traffic and six thousand samples of normal traffic in the SDSN dataset to test external attacks. We integrated these samples with the eight thousand MANSIM data to build a comprehensive smart seaport dataset for external attacks.
The depiction of performance comparison of external attacks can be seen in Fig.~\ref{fig:external}. 
The results showed that our solution performed better than DNN for internal and external attacks.

We also sent a specific number of internal and external attacks to the system simultaneously to scrutinize the system's performance. Then, we compared the detection rate of our PS and the DNN algorithm. We defined the detection rate between 0 and 1. ``0'' indicates out-of-detection,  and ``1'' means that all attacks are detected. 
We examined the system's simultaneous attack performance for all combinations of six internal and six external attacks.
Fig.~\ref{fig:detrate} depicts the detection rate of simultaneous attacks.
The result shows that the detection performance of our solution is better and more robust than DNN. Overall, our solution successfully detects both internal and external attacks.

\section{Conclusion}
\label{sec:conc}

In this study, we design a DT-assisted honeypot, TwinPot, for next-generation smart seaports.
We divide the seaport network traffic into internal and external traffic in the proposed system. The traffic from the existing assets of the seaport is what we call internal traffic. We refer to external traffic as traffic from entities other than existing entities in the seaport, such as IoT devices on inbound cargo ships.
We also propose an intelligent attack detection mechanism to handle different attack types using DT technology for the current entities in seaports. Lastly, we test the performance of our solution using the MANSIM tool and two existing datasets to build an extensive smart seaport dataset for internal and external attacks. We compare our solution with the DNN algorithm. 
The results showed that our solution performed better than DNN for internal and external attacks.
We also show that under both simultaneous internal and external attacks on the system, our solution successfully detects internal and external attacks.

\section*{Acknowledgment}
Yagmur Yigit would like to thank the ITU-Turkcell and ITU-BTS Graduate Research Scholarship Programme for their support.

\bibliographystyle{IEEEtran}



\end{document}